\documentclass{article}
\usepackage{emulateapj,apjfonts}

\def\kms{km s$^{-1}$}

\def\mh{{M_{\bullet}}} 
 
\def\msun{{M_{\odot}}}
 
\def\ms{{\mh-\sigma_c}} 
\def\mm{{\mh-M_{DM}}}
 
\def\vv{{\textsl v_{vir}}} 
\def\vc{{\textsl v_c}}
 
\def\vs{{\vc-\sigma_c}} 
\def\ml{{\mh-L_B}}
\def\lae{\mathrel{<\kern-1.0em\lower0.9ex\hbox{$\sim$}}}
\def\gae{\mathrel{>\kern-1.0em\lower0.9ex\hbox{$\sim$}}}

\lefthead{Ferrarese} \righthead{Supermassive Black Holes and Dark
Matter Halos}

\begin{document}

\title{Beyond the Bulge: A Fundamental Relation Between Supermassive
Black Holes and Dark Matter Halos}

\author{Laura Ferrarese} 
\affil{Rutgers University, New Brunswick, NJ,
08854,  lff@physics.rutgers.edu} 
\authoraddr{Department of Physics and
Astronomy, 136 Frelinghuysen Road,  Piscataway, NJ 08854}

\begin{abstract}
The possibility that the masses $\mh$ of supermassive black holes
(SBHs) correlate with the total gravitational mass of their host
galaxy, or the mass $M_{DM}$ of the dark matter halo in which they
presumably formed, is investigated using a sample of 16 spiral and 20
elliptical galaxies.  The bulge velocity dispersion $\sigma_c$,
typically defined within an aperture of size $R \lae 0.5$ kpc, is
found to correlate tightly with the galaxy's circular velocity $\vc$,
the latter  measured at distances from the galactic center at which
the rotation curve is flat,  $R \sim 20-80$ kpc. By using the well
known $\ms$ relation for SBHs, and a prescription to relate $\vc$ to
the mass of the dark matter halo $M_{DM}$ in a standard $\Lambda$CDM cosmology,
the correlation between $\sigma_c$ and $\vc$ is equivalent to one
between $\mh$ and $M_{DM}$. Such a correlation is found to be
nonlinear, with the ratio $\mh/M_{DM}$ decreasing from
$2\times10^{-4}$ for $M_{DM} \sim  10^{14}$ $\msun$, to $10^{-5}$ for
$M_{DM} \sim 10^{12}$ $\msun$. Preliminary evidence suggests that
halos of mass smaller than $\sim 5\times 10^{11}$ $\msun$ are
increasingly less efficient -- perhaps unable -- at forming SBHs.
\end{abstract}

\section{Introduction}
Never has the study of supermassive black holes (SBHs) been greeted
with such attentiveness as in the past few years.  Although by the mid
1960s it was generally recognized that the energy source of the newly
discovered quasars must be gravitational (e.g. Robinson 1965), for the
following three decades the existence of SBHs  was destined to be
surrounded by skepticism. By the mid 1990s, a few years after the
launch of the Hubble Space Telescope, it was widely accepted (Kormendy
\& Richstone 1995).  Today,  SBHs feature prominently in models for
the formation and evolution of their host galaxies (e.g. Silk \& Rees
1998). Freed from the burden of having to demonstrate the very
existence of SBHs, we can now begin asking cardinal questions: how did
SBHs  form, how do they accrete, how do they evolve, and what role  do
they play in the formation of cosmic structure?

Tackling such questions has been made possible by the discovery  that
SBHs are tightly related to the large-scale properties of their host
galaxies.  Black holes masses, $\mh$, correlate with the blue
luminosity $L_B$ of the host bulge (Kormendy \& Richstone 1995). Even
more compelling is the correlation between $\mh$ and the bulge
velocity dispersion, $\mh \propto \sigma^{\alpha}$ (Ferrarese \&
Merritt 2000; Gebhardt et al. 2000). Using the most up-to-date set of
SBH mass measurements such a correlation can be written as:

\begin{equation}\mh = (1.66 \pm 0.32) \times 10^8 (\sigma_c/200~ {\rm 
km~s}^{-1})^{4.58 \pm 0.52}~~\msun\end{equation}

\noindent (Ferrarese 2002), where $\sigma_c$ is normalized to an aperture of size one
eighth the bulge effective radius. The $\ms$ relation has supplanted
the $\ml$ relation as the tool of choice for the study of SBH
demographics (Ferrarese 2002; Yu \& Tremaine 2002), for the following
reasons. The $\ml$ relation suffers from a large  scatter (a factor
of several in $\mh$, Ferrarese \& Merritt 2000),  most (but not all)
of which is ascribable to lenticular and spiral galaxies (McLure \&
Dunlop 2002).  The $\ms$ relation, on the other hand, appears to be
intrinsically  perfect (Ferrarese \& Merritt 2000; Merritt \&
Ferrarese 2001a; Ferrarese 2002). It holds true for galaxies of
disparate Hubble types (from SBs to compact ellipticals to cDs)
belonging to wildly different environments (from rich clusters to the
field), showing perfectly smooth (e.g. NGC 6251) or highly disturbed
(e.g. NGC 5128) morphologies. Furthermore, $\sigma$ is a more
faithful gauge of mass than $L_B$: the $\ms$ relation thus betrays the
existence of a tight connection between the masses of SBHs and those
of the hot stellar component surrounding them (Merritt \& Ferrarese
2001b).

Have we indeed found, in the $\ms$ relation, the fossil record that
links the formation and evolution of SBHs to those of their host
galaxies? The $\ms$ relation, and the equally tight relation recently
discovered by Graham et al. (2001) between $\mh$ and the central
concentration of bulge light,  probe the connection between SBHs and
the baryonic component of galactic bulges. Measurements of $\sigma$
typically do not extend far enough to penetrate the region dominated
by the dark matter (DM) component (Faber \& Gallagher 1979).  And yet,
it is not the mass of the bulge but rather, the {\it total mass}
$M_{tot}$  of the galaxy or the mass $M_{DM}$ of the DM halo, which is
the key ingredient in most theoretical models proposed for the
formation of SBHs (Adams, Graff \& Richstone 2000; Monaco et al. 2000;
Haehnelt, Natarajan \& Rees 1998; Silk \& Rees 1998; Haehnelt \&
Kauffmann 2000; Cattaneo, Haehnelt \& Rees 1999; Loeb \& Rasio
1994). Once a correlation with $M_{tot}$ (or $M_{DM}$) is recovered by
the models, the correlation with bulge mass is implicit because, in
standard CDM scenarios, the bulge mass is loosely determined by the
halo properties (e.g. van den Bosch 2000; Haehnelt, Natarajan \& Rees
1998; Zhang \& Wyse 2000).

This paper pursues, from an observational standpoint, the possibility
of a link between $\mh$, $M_{tot}$ and $M_{DM}$.   A tight correlation
is indeed found in the form of a relation between $\mh$ and the
circular velocity of the SBH host galaxy, measured well beyond the
optical radius.

\section{Database}

Measuring the total gravitational mass of a galaxy is a formidable
challenge.  In bright ellipticals, the temperature and density
distribution of the thermal  X-ray gas can lead to a measure of
$M_{tot}$ under the minimal condition of hydrostatic equilibrium
(e.g. Fabricant, Rybicki \& Gorenstein 1984). However, while much
progress is expected as Chandra data are collected, 

\centerline{\includegraphics[width=8.0cm]{f1.epsi}}
\figcaption{Correlation between bulge velocity dispersion $\sigma_c$ and
disk circular velocity $\vc$ for a sample of 37 galaxies with either
optical (open circles) or HI (solid circles) rotation curves. Points
marked by errorbars only correspond to galaxies for which the rotation
curves does not extend beyond $R_{25}$. The galaxy to the far left,
with the smallest value of $\sigma_c$, is NGC 598. The solid line
corresponds to a fit to all galaxies with $\sigma_c > 70$ \kms~and
$R(\vc)/R_{25} > 1.0$. The dotted line corresponds to the fit to all
galaxies, with the only exclusion of NGC 598.}  
\centerline{}
\vskip .2in

\noindent previous generations of X-ray satellites lacked the
necessary combination of spectral response and sensitivity to allow an
accurate measurement of all  required observables (e.g. Biermann et
al. 1989; Awaki et al. 1994; Matsushita et al. 1998). Stars, globular
cluster and planetary nebulae (PNe) are good tracers of  the
gravitational potential (e.g. Saglia et al. 2000; C\^ot\'e et
al. 2001; Arnaboldi et al. 1998). However, the derived masses suffer
large uncertainties  due to the unknown velocity anisotropy of the
system and, in particular for the case of PNe, the small sample size
of the tracers (Napolitano et al. 2001; Merritt \& Tremblay 1993).

In spiral galaxies, however, the rotation curves of the cold disk
component bear the direct imprint of the mass distribution.  Indeed,
the lack of a decline in the circular velocity $\vc$ at radii of order
and beyond the optical radius of spirals was one of the first
indications of the existence of a significant dark matter component
(Rubin, Ford \& Thonnard 1978, 1980; Rubin, Thonnard \& Ford 1977;
Krumm \& Salpeter 1979). To first order, therefore, a relation between
$\mh$ and $M_{tot}$ would be reflected in a relation between $\mh$ and
$\vc$.

Of the four spiral galaxies with a secure measurement of $\mh$ (see
the compilation by Merritt \& Ferrarese 2002) only two, the Milky Way
and NGC 4258, have published rotation curves, both derived from 21-cm
studies  (Olling \& Merrifield 1998;  van Albada 1980). However, the
sample  can be augmented if one is willing to exchange $\mh$ for the
bulge velocity dispersion $\sigma_c$, following equation (1). Of all
spirals for which $\sigma$ has been measured, 46 have rotation curves,
either in HI (21 objects) or optical (25 objects).  For six of the
galaxies with an HI rotation curve, warnings are found in the original
papers as to the use of the published  circular velocities for mass
measurements. Reasons include warps in the HI disk preventing one from
deprojecting  the observed rotation curve (NGC 4565, Rupen 1991; NGC
891, Rogstad \& Shostak 1972; NGC 1808, Saika et al.  

\centerline{\includegraphics[width=8.0cm]{f2.epsi}}
\figcaption{Residuals from the best fit to all galaxies with the
exclusion of NGC 598 (dotted line in Fig. 1), plotted as a function of
$R(\vc)/R_{25}$.}
\centerline{}
\vskip .2in

\noindent 1980, Koribalski
et al. 1993) and/or the presence of significant non-circular motion in
the HI gas (NGC 3079, Irwin \& Seaquist 1991; NGC 3031, Rots \& Shane
1975; NGC 4736, van der Kruit 1974). For two of the galaxies with an
optical rotation curve (NGC 3200 and NGC 7171), two values of $\sigma$
discrepant at the $> 99.9$\% level exist in the literature (McElroy
1995; Whitmore et al. 1984; Dressler 1984);  these galaxies have also
been excluded from the sample.

Data and references for the remaining 38 galaxies are given in Table
1.  To allow for the application of equation (1), the bulge velocity
dispersion $\sigma$ is corrected to the equivalent of an aperture of
radius  $r_e/8$ following the prescriptions of Jorgensen, Franx \&
Kjaergaard (1995). The bulge effective radius  $r_e$ is taken from
Baggett et al. (1998, but see notes to Table 1).

\section{The $\vs$ Relation}

Fig. 1 shows the circular velocity versus the bulge velocity
dispersion within $r_e/8$ for all galaxies in Table 1. The correlation
is immediately apparent, but a few observations are in order.

First, a glance at Table 1 will reveal that the most noticeable
difference between the two samples of optical and HI rotation curves
is, not surprisingly, that while only two of the  optical rotation
curves extend beyond $R_{25}$, the galaxy radius at the $B=25$ mag
arcsec$^{-2}$ isophote, all but one of the HI rotation curves do.
Inspection of the original papers shows that the HI rotation curves
all  exhibit a common pattern. It is well known (e.g. Sofue \& Rubin
2001 and references therein) that  as one moves away from the center,
$\vc$ rises rapidly, peaks somewhere between 3 and 15 kpc, and then
remains flat (sometimes after a  modest initial decline) until the
last measured point, typically $R(\vc) = 30-40$ kpc.   Optical
rotation curves rarely extend beyond 10 kpc; at this distance $\vc$
has not yet settled into a flat rotation curve.  In Fig. 1, galaxies
with rotation curves not extending beyond $R_{25}$ (shown as errorbars
only) appear to display larger scatter, relative to the mean
correlation defined by all galaxies, than the galaxies with a more
extended rotation curve.

\centerline{\includegraphics[width=8.0cm]{f3.epsi}}
\figcaption{Same as Fig. 1, except solid dots represent spiral galaxies
with $R(\vc)/R_{25} > 1.0$, open dots represent the sample of
elliptical galaxies from Kronawitter et al. (2000). The dotted line is
a fit to all spiral galaxies with the exception of NGC 598, while the
solid line is the fit to all spirals with $\sigma_c > 70$ \kms. The
latter fit is consistent with the one obtained for the elliptical
galaxies alone (see fits 3 and 4 in Table 2). The dashed line is 
a fit to the entire sample of ellipticals plus spirals with $\sigma_c
> 70$ \kms.}\centerline{}
\vskip .2in

Second, a simple visual inspection of Fig. 1 shows that among the
galaxies with $R(\vc)/R_{25} > 1$, NGC 598 (= M33, the galaxy with the
lowest value of $\sigma_c$ in the plot) is a clear
outlier. Furthermore, the two galaxies with $\sigma_c < 70$ \kms~seem
to have a significantly larger circular velocity that would be
inferred  if a linear fit (in $\log\sigma_c - \log\vc$) were to be
performed to the rest of the sample.

To quantify the above statements, we performed a regression analysis
(Akritas \& Bershady 1996)
to all galaxies except NGC 598 (Table 2 and  dotted line in
Fig. 1). Fig. 2 shows the residuals from the best fit as a function of
$R(\vc)/R_{25}$, confirming the larger scatter for galaxies with
rotation curves not extending beyond $R_{25}$. For the scope of this
paper, the usefulness of $\vc$ is directly related to how confidently
$\vc$ can be used as an estimator of the total galactic mass, which in
turns depends critically on the radial extent of the rotation curve.
Therefore, after dutifully noting that the regression analysis
produces consistent fits whether the entire sample, or just
the galaxies for which $R(\vc)/R_{25} > 1$ are used (Table 2), we will
only retain the latter sample in the remainder of this paper.

Because of its apparently negligible intrinsic scatter  (a reduced
$\chi^2_r$ of 0.70, Table 2) and the fact that it does not follow
directly from simple dynamical arguments (see \S 3.1), the $\vs$
relation is of obvious interest. However, by targeting only spiral
galaxies it lacks generality; furthermore, it samples a narrow range
in $\sigma_c$, $\sim 90$ to $\sim 180$ \kms. It is interesting to
compare the $\vs$ relation observed for the spirals with the one
derived recently by Kronawitter et al. (2000) and Gerhard et
al. (2001) for a sample of 21 elliptical galaxies. For this sample,
circular velocity curves are derived by applying non-parametric
spherical models to the observed stellar absorption line profiles,
velocity dispersion and surface brightness profiles.  Kronawitter et
al.  conclude that the rotation curves resulting from the models are
very likely nearly self-similar: all rotation curves are consistent
with being flat outside 0.3$r_e$, with $r_e$ the galaxy's effective
radius\footnotemark.  As pointed out by Gerhard et al. (2001), $\vc$
is expected to be linearly related to any other velocity scale, for
instance $\sigma_c$, as a consequence of the near dynamical homology.  The
constant of proportionality is likely to depend slightly on the
details of the models; Kronawitter et al. (2000) and Gerhard et
al. (2001) do not discuss this point explicitely, but by comparing
their results to the those obtained by van der Marel (1991) and Lauer
(1985) for the same galaxies, variations of the order of a few tens of
a percent are possible if different dynamical models are used.

\footnotetext{NGC4486B is the only galaxy for which the models are
deemed unreliable, and will not be considered further.}

Fig. 3 shows the $\vs$ relation for the combined samples of 16 spirals
with $R(\vc)/R_{25} > 1$ (solid circles) plus the 20 ellipticals from
Kronawitter et al. (2000, open circles). Values of
$\sigma$ for the ellipticals are taken from Davies et al. (1987), and
corrected to a common aperture size as detailed for the bulges. As
expected, we recover the correlation between $\vc$ and $\sigma_c$
already reported by Gerhard et al. (2001).

The fact that the spiral  and elliptical samples follow the same
relation is unexpected (see \S 3.1) and points to a common cause
connecting galaxy dynamics at small and large scales across the entire
Hubble sequence. A regression analysis to the ellipticals produces a
fit consistent with the one obtained for the spirals once the two
galaxies with $\sigma_c < 70$ \kms~are excluded (fits 3 and 4 in Table
2). If a linear fit is forced, $\sigma_c = (0.65 \pm 0.02) \vc $ for
the ellipticals and $\sigma_c = (0.61 \pm 0.01) \vc$ for the spirals.
However, because the correlation between $\vv$ and $\sigma_c$ for the
ellipticals is partly due to the scale-free nature of the models,
we use only the sample of 13 spirals with $R(\vc)/R_{25} >
1$ and $\sigma_c > 70$ \kms~ in deriving the best fit to the $\vs$ relation:

\begin{equation}\log{\vc} = (0.84 \pm 0.09) \log{\sigma_c} + (0.55\pm0.19)\end{equation}

While this fit is strictly derived in the  $70 < \sigma_c < 180$
\kms~range, it can also be considered valid in the $120 < \sigma_c <
350$ \kms~range populated by the ellipticals (Fig. 3 and Table 2).  As is
the case for the $\ms$ relation, the reduced $\chi^2_r = 0.38$ of the
fit points to a relation  with negligible intrinsic scatter.

\subsection{Is the $\vs$ Relation a Tautology?}

The tightness of the $\vs$ relation is suprising to the point
of arising suspicions that we might be dealing with a tautology.

The most trivial case to consider is the following: a strict
proportionality between $\sigma_c$ and $\vc$ would result if both
responded to the same mass distribution. For instance,  if the region
within which $\sigma_c$ is measured extends to radii where the
rotation curve is flat, $\sigma_c$ would naturally be linearly related
to $\vc$ simply by virtue of the virial theorem. This is not the case
for any of the galaxies considered here: even for the bulges with the
largest effective radius, $\sigma_c$ only samples the innermost 0.5
kpc, while the rotation curve does not settle to a constant velocity
until much farther out, typically 4 to 20 kpc. Fig. 4 plots the ratio
of bulge central velocity dispersion to disk circular velocity against
the ratio of the radii at which the two are measured (the radius at
which  $\vc$ peaks is taken as a lower limit to the radius at which
the  circular velocity first settles onto a flat curve). As can be
seen, $R(v_{max})/R(\sigma_c)$  ranges from values of a few to over
100.

Even so, if rotation curves of spiral galaxies formed a nearly
homologous family, as is the case for the elliptical galaxies of
Gerhard et al. (2001),  $\vc$ or $\sigma_c$ would be linearly related

\centerline{\includegraphics[width=8.0cm]{f4.epsi}}
\figcaption{The ratio between bulge velocity dispersion $\sigma_c$ and
disk circular velocity $\vc$ plotted against the ratio of the radii at
which the two are measured. Solid circles are galaxies for which
$R(\vc)/R_{25} > 1.0$.}\centerline{}
\vskip .2in

\noindent(however, the constant of proportionality would depend on the details
of the mass density profile, and there would be no reason for the
elliptical and spiral samples to define a common relation). Unlike
Gerhard's ellipticals, however, spiral galaxies show a strong
dynamical non-homology: both the profile shape and the amplitude of
the rotation curves depend on the galaxy luminosity (Casertano \& van
Gorkom 1991; Persic, Salucci \& Stel 1996). Therefore, a linear
relation, or a relation at all, between $\vc$ and $\sigma_c$ is not
expected a priori.

The two arguments presented above also explain why the $\vs$ relation
is not  a direct result  of the so called ``disk-halo
conspiracy''. Most rotation curves are characterized by a reasonably
flat rotation curve in the outer parts. This implies   that the mass
density profiles of the luminous and dark matter must be coupled,
thereby ``conspiring'' to produce a flat rotation curve and the
absence of conspicuous features marking the edge of the luminous
component (Casertano \& van Gorkom 1991). However  the conspiracy
breaks down in the inner, bulge dominated region, which is always
characterized by a steeply increasing rotation curve whose detailed
shape and overall slope depends strongly on the galaxy luminosity  and
morphological type (Persic, Salucci \& Stel 1996; Sofue et al. 1999).
For the galaxies in our sample, only for NGC 2841 the case could be
made of a bulge-disk conspiracy: the rotation curve is approximatively
flat all the way to the innermost point, $\sim 2$ kpc (Begeman
1987). In all other cases, the bulge contribution is insignificant by
the time the flat part of the rotation curve is reached (Broeils 1992;
Kent 1987). In other words there is no direct evidence in these
galaxies of a coupling between bulge and halo. At most, a connection
between the $\vs$ relation and the disk-halo conspiracy can be thought
of in the following terms. A rough correlation exists between maximum
rotational velocity and Hubble type (e.g. Casertano \& van Gorkom
1991); coupled with the correlation between Hubble type and bulge
luminosity and between the latter and the bulge velocity dispersion
(Kormendy \& Illingworth 1983), a correlation between $\vc$ and
$\sigma_c$ ensues. However, as in the case for the $\ms$ relation,
which is ``expected'' given the $\ml$ and the Faber-Jackson relation,
the $\vs$ relation is tighter, and therefore more fundamental, than
any of the correlations mentioned above.

One last possibility to explore is whether  the $\vs$ relation could
be nothing but the Tully-Fisher relation in disguise. Verheijen (2001)
finds negligible intrinsic scatter for the $K$-band Tully-Fisher
relation when the circular velocity in the flat part of the rotation
curve is substituted to the maximum rotational velocity. The tightness
of the relation implies a fundamental connection between DM halo mass
and the total baryonic mass.  The connection with bulge mass (and
hence $\sigma_c$) is however  not immediate (Norman, Sellwood \& Hasan
1996). For instance, the bulge to disk fraction could conceivably
depend on the detailed form of the halo angular momentum profile
(Bullock et al. 2001) and the details of the angular momentum exchange
between gas clouds during dissipation (van den Bosch et al. 2002). It
should be further noticed that Verheijen (2001) finds that the
Tully-Fisher relation holds with negligible intrinsic scatter down to
$\vc \sim 80$ \kms, while the $\vs$ relation shown in Figure 3 seems
to break down below $\vc \sim 150$ \kms.

\section{A Relation Between SBHs and DM Halos}

Not being the result of a dynamical tautology, the near invariance of
$\sigma_c/\vc$ over two orders of magnitude in $R(\vc)/R(\sigma_c)$
seems to indicate a remarkable uniformity in the (luminous + dark)
mass density profile along the Hubble sequence.   In particular,
characterizing the slope and normalization of the relation might help
in constraining theoretical models and numerical simulation
(e.g. Steinmetz \& Muller 1995) following the formation and evolution
of galaxies.

Furthermore, as discussed in \S 2, the existence of the $\vs$ relation
is strongly suggestive of a tight correlation between $\mh$ and the
total gravitational mass of the SBH host galaxy.  If the rotation
curves for all galaxies  plotted in Fig. 3 are flat at a distance  $R$
from the center,  i.e. ${\textsl v}(R) = \vc$, the virial theorem
ensures that $M(R) \propto \vc^2$.  Equations (1) and (2) then imply
$\mh \propto M(R)^{2.7}$.

Several proposed scenarios trace the origin of SBHs to the very early
stages of structure formation (Cattaneo, Haehnelt \& Rees 1999; Silk
\& Rees 1998; Haehnelt \& Rees 1993; Umemura, Loeb \& Turner 1993). By
defining the depth of the potential well, the mass and distribution of
the DM halo control not only the formation of SBHs, but also their
relationship to the luminous matter. Relating $\vc$ to $M_{DM}$ is no
trivial task, but at least a rough attempt can be made as follows. In
a CDM-dominated universe, $M_{DM}$ is uniquely determined by the halo
velocity ${\textsl v}_{vir}$ measured at the virial radius,
$R_{vir}$. The latter is defined as the radius at which  the mean
density exceeds the mean universal density by a constant factor,
generally referred to as the ``virial overdensity'' $\Delta_{vir}$.
Based on this definition, and in virtue of the virial theorem, it
immediately follows that  $M_{DM}$ must be proportional to the third
power  of ${\textsl v}_{vir}$. Being a function of $\Delta_{vir}$, the
constant of proportionality depends on the adopted cosmology and may
vary with time (e.g. Bryan \& Norman 1998; Navarro \& Steinmetz 2000).
For the purpose of this paper, we will adopt the same $\Lambda$CDM 
cosmological model used by Bullock et al. (2001)\footnotemark. 
Under these assumptions:

\footnotetext{$\Omega_{\Lambda}=0.7$, $\Omega_m=0.3$, $h = 0.7$}

\begin{equation}
M_{DM} = 2.7\times10^{12} (\vv/200~ {\textrm \kms})^3~ \msun
\end{equation}

\centerline{\includegraphics[width=8.0cm]{f5.epsi}}
\figcaption{Same as Fig. 3, but with $\sigma_c$ transformed into SBH mass
using equation (1), and $\vc$ transformed into DM halo mass following
the prescriptions of Bullock et
al. (2001).  Solid circles are spirals, solid triangles
ellipticals. Symbols accentuated by a larger open contour identify
galaxies having a dynamical estimate of $\mh$, which was used in the
plot (Merritt \& Ferrarese 2002). The upper limit on the SBH mass in
NGC 598 is marked by the arrow. The dotted line represents a constant
ratio  $M_{DM}/\mh = 10^{5}$. The solid line corresponds to the
best fit using Bullock's et al. prescription to relate $\vv$ to $\vc$
(equation 5). The dashed line shows where the galaxies would lie if
 $\vv = \vc$ (equation 4), while if $\vc/\vv =1.8$, as proposed by Seljak
(2002), the points would move to the location shown by the dot-dashed line
(equation 7). }\centerline{}
\vskip .2in

\noindent (Bullock et al. 2001). 

In a zeroth order approximation, we might assume that $\vv \sim \vc$. 
Combining equations (1), (2) and (3) gives:

\begin{equation}{{\mh} \over {10^8~\msun}} \sim 0.027 {\left({M_{DM}} 
\over {10^{12}~ \msun}\right)}^{1.82}\end{equation}

However,  for any reasonable shape of the DM density profile, the
circular velocity will decrease towards $R_{vir}$, in other words,
$\vc$ as listed in Table 1 is most certainly an upper limit to $\vv$.
The shape of the rotation curve, and therefore the $\vc/\vv$ ratio is 
of course strongly dependent on the 
assumed DM density profile. We will again follow Bullock et al. (2001)
in adopting an NFW profile (Navarro, Frenk \& White 1995)
characterized by an ``inner'' radius $r_s$. In this specific
application, $\vv/\vc(R)$ is a function of two variables, the
concentration parameter $C_{vir} = R_{vir}/r_s$ and the ratio
$R/r_s$ (Bullock et al. 2001). N-body simulations (Bullock et al. 2001) show that for
present day halos in the $10^{11} - 10^{14}$ $\msun$ range, $C_{vir}$
depends weakly on $M_{DM}$: $C_{vir} \sim 5\times10^2
M_{vir}^{-0.13}$, where $M_{vir}$ is in units of $\msun$. Therefore,
for $R_{vir} \sim {\rm a~few}\times10^2$ kpc (Klypin et al. 2001), $r_s =
R_{vir}/C_{vir} \sim {\rm a~few~to~a~few}\times10$ kpc, i.e. $r_s$ is
likely sampled by the flat HI rotation curves used in deriving the
$\vs$ relation. This is a reasonable conclusion: in the region of
space dominated by DM, the turnover in the rotation curve can be
roughly expected to coincide with the turnover in the DM density
profile. Therefore, by assuming $R = r_s$, $\vv/\vc(R)$  is a function
of $M_{DM}$ only. By using equation (3) to 
express $\vv$ as a function of $M_{DM}$, Bullock's et al. results
can be approximated as:

\begin{equation}{{M_{DM}} \over {10^{12}~\msun}} \sim 1.40 {\left({\vc} 
\over {200~ {\rm km~s}^{-1}}\right)}^{3.32}.\end{equation}

By using equation (5) to transform $\vc$ to $M_{DM}$, and equation (1)
to transform $\sigma_c$ to $\mh$, equation (2) can be  written as
(Fig. 5):

\begin{equation}{{\mh} \over {10^8~\msun}} \sim 0.10 {\left({M_{DM}} 
\over {10^{12}~ \msun}\right)}^{1.65}\end{equation}

A third, completely independent way of converting the $\vs$ relation into a
relation between  SBH and DM halo masses is the following.
Observational constraints on the mass density profiles of DM halos have been
recently derived (Seljak 2002) by combining observations of
galaxy-galaxy lensing with the Tully-Fisher and fundamental plane
relations.  For $L_*$ galaxies, the circular velocity at the virial
radius decreases  by a factor 1.8 from its peak value at the optical
radius. This decline is steeper than predicted by Bullock's et
al. (2001) numerical simulations, possibly due to the neglect, in the
latter work, of the baryonic contribution to the rotation curves
at optical radii (Seljak 2002).

By assuming $\vv/\vc(R) \sim 0.56$\footnotemark, the relation between $M_{DM}$ and
$\mh$ becomes:

\begin{equation}{{\mh} \over {10^8~\msun}} \sim 0.67 {\left({M_{DM}} 
\over {10^{12}~ \msun}\right)}^{1.82}\end{equation}

\footnotetext{This is likely a lower limit to $\vv/\vc(R)$, since
the peak circular velocity is $\ge \vc(R)$.}

Comparing  equations (4), (6) and (7)  can give a rough idea of how
the functional dependence of $M_{DM}$ on $\mh$  depends on the
assumptions made for the DM density profile and the resulting rotation
curve.  

\subsection{The Significance and Scatter of the $\mm$ Relation}

The $\mm$ relation implies that the formation of SBHs is controlled,
perhaps indirectly, by the properties of the DM halos in which they
reside. Whether the connection between SBHs and DM halos is of a more
fundamental nature than the one between SBHs and bulges,  reflected in
the $\ms$ relation, is unclear at this time.  The $\ms$ relation
implies a scatter of the order of $20-25$\% in $\mh$ for $\sigma_c = 70
- 300$ \kms. The scatter in the $\mm$ relation is  difficult to
determine, since the uncertainty in the transformation between $\vc$
and $M_{DM}$ is largely unknown. Bullock et al. (2001) estimate
$\Delta \log(C_{vir}) = 0.18$. In the range $\vc = 50 - 400$ \kms,
this translates into a 5\% to 7\% error in $M_{DM}$ in equation (5),
and a 9\% to 12\% error on $\mh$ in equation (6). This indicates that,
at least formally, the scatter in the $\mm$ relation  could be a
factor two smaller than the scatter in the $\ms$ relation for any
given choice of DM density profile and cosmological parameters.
Secure conclusions, however, will have to await an empirical
characterization of the $\mm$ relation, with both $\mh$ and $M_{DM}$
determined directly from observations.

Additional clues on how bulges and DM halos influence the 
formation of SBHs can be derived by studying bulgeless
spirals. Such galaxies  have been forcefully neglected in the present
study, where we required the existence of a bulge velocity dispersion
in order to derive $\mh$.  Of the galaxies in Table 1, NGC598 is the one with the
least conspicuous bulge\footnotemark. IC342 and NGC 6503 are late type spirals
with minimal bulges; like NGC598, they deviate significantly from the
$\vc$ relation defined by the more massive systems. All of the
galaxies with no or minimal  bulge for which  a rotation curve exists,
including both dwarf systems and low surface brightness (LSB)
galaxies,  have circular velocities below 150 \kms
(e.g. Blais-Oulette, Amram \& Carignan  2001; de Blok \& Bosma
2002). In particular, in the LSB compilation of  van den Bosch et
al. (2001) the only two galaxies with maximum circular velocity in
excess of 150 \kms~ are also the only galaxies with a significant bulge
component. Therefore, it does seem that bulges, as
well as SBHs, are always formed in halos with circular velocity in
excess of $\sim 150$ \kms. In less massive halos, the interplay
between halo, bulge and the central SBH remains to be
investigated. For this purpose attempts to directly measure the mass
of a putative SBH in bulgeless galaxies is of primary importance.

\footnotetext{The existence of a rarefied bulge in
this galaxy is supported by both spectroscopic and photometric data
(e.g. Minniti, Olszewki, \& Rieke 1993; Regan \& Vogel 1994; Mighell
\& Rich 1995; Minniti 1996).}

Whether equation (4), (6) or (7) is adopted,  $\mh$ does not depend
linearly on $M_{DM}$.  According to equation (6), the ratio between
SBH and halo mass decreases from $\sim 2\times10^{-4}$ at $M_{DM} =
10^{14}$ $\msun$ to $\sim 10^{-5}$ at  $M_{DM} = 10^{12}$ $\msun$,
i.e. less massive halos are less efficient in forming SBHs.  
This tendency becomes more pronounced for halos with $M_{DM} \lae
5\times10^{11}$ $\msun$: as mentioned above, NGC 6503, NGC 3198 and NGC 598 lie above the
relation defined by the more massive galaxies. It must be pointed out,
however, that the $\ms$ relation is not defined for $\mh < 10^6$
$\msun$, indeed we have no direct evidence of the existence of nuclear
black holes in this mass range.  If such SBHs exist, the recent report
of an upper limit of only a few $10^3$ $\msun$ for the BH  in NGC 598
(Merritt, Ferrarese \& Joseph 2001; Gebhardt et al. 2001), indicates
that they are not more massive than  would be inferred by
extrapolating the $\ms$ relation to the $\sigma_c \sim 20$ \kms~
range.  This excludes the possibility that NGC 6503 and NGC 3198 could
be reconciled with the $\mm$ relation defined by the more massive
galaxies. 

These findings are in qualitative agreement with the formation
scenario envisioned by Haehnelt, Natarajan \& Rees (1998) and Silk \&
Rees (1998).  In an attempt to match the DM halo mass function to the
luminosity function of quasars, Haehnelt et al.  noted that as the
quasar lifetime is increased, a match can be obtained only at the
expense of steepening the slope of the relation between $\mh$ and
$M_{DM}$. In their model,  a linear relation between $\mh$ and
$M_{DM}$ implies a quasar lifetime $t_{QSO} \sim 3 \times 10^5$ yr,
about a factor 100 shorter than favored by current observational
constraints (e.g. Martini \& Weinberg 2001; Ferrarese 2002).  If
$t_{QSO}\sim {\rm a~few}\times 10^7$ yr,  the slope of the $\mm$
relation must be increased to $\sim 1.7$, in rather remarkable
agreement with the results derived in this paper.

Furthermore, Haehnelt et al. propose that SBH formation takes place
during the very early stages of galaxy formation.  Gas gravitating in
a DM halo is deposited at the center of the halo at a rate which is
proportional to the third power of the halo circular velocity
$\vv$. Stars more massive than $\sim 10^6$ $\msun$ undergo
post-Newtonian instabilities -- and therefore collapse -- before they
can ignite hydrogen; for $\vv$ larger than $\sim 200$ \kms, such stars
can be formed on short enough timescales that the collapse will take
place before the star has had a chance to burn all of its fuel. Only
halos with $\vv \gae 200$ \kms~(or masses $\gae 10^{12}$ $\msun$) would
therefore host SBHs, and only SBHs with $\mh > 10^6$ $\msun$ can be
formed in such halos. It is therefore possible that galaxies such as 
NGC 6503 and NGC 3198 do not contain SBHs.

\section{Conclusions}

The results in this paper can be summarized as follows:

\begin{itemize}

\item{We have considered a sample of 15 spiral galaxies with published
measurements of both the bulge velocity dispersion $\sigma_c$ and the
disk circular velocity $\vc$, measured beyond the optical radius.  The
sample displays a tight correlation between the two quantities.
Furthermore, in a $\log(\sigma) - \log(\vc)$ plane, the narrow strip
occupied by the spirals blends smoothly with that occupied by a sample
of 20 elliptical galaxies (from Kronawitter et al. 2001 and Gerhard et
al. 2001). This is an interesting result, in view of the different
methods used to derive $\vc$ for the two samples, and of the
dissimilarities between the shapes of the rotation curves observed in
spirals and modelled in elliptical galaxies.}

\item{The ellipticals and spirals define a common, linear (in a
log-log plane) $\vs$ relation for $70 < \sigma_c < 350$ \kms. Three
galaxies have $\sigma_c$ below this range; all three deviate from the
relation in the sense of having smaller than expected values of
$\sigma_c$. The discrepancy is largest -- a factor 7 larger than the
standard deviation on $\sigma_c$ -- for the galaxy with the smallest
value of $\sigma_c$, NGC 598.}

\item{Rotation curves for all of the galaxies appear to be flat in the
outer parts. In virtue of the $\ms$ relation and the virial theorem,
the $\vs$ relation must therefore imply that  the masses of
supermassive black holes correlate with the gravitational mass of
their host galaxy, measured within a given radius $R > R_{25}$.}

\item{If the further concession is made that, in CDM scenarios of
structure  formation, the virial mass of the DM halo is uniquely
determined by $\vc$ (Navarro \& Steinmetz 2000, Bullock et al. 2001),
the $\vs$ relation is the observational evidence that SBH masses
correlate tightly with the mass of the DM halos in which, presumably,
they are formed. The relation between $\mh$ and $M_{DM}$ is nonlinear:
the ratio between $\mh$ and $M_{DM}$  decreases with decreasing halo
mass. This trend becomes more severe for $M_{DM} \lae 5\times 10^{11}$
$\msun$. It is possible that less massive halos might indeed be unable
to form SBHs, in qualitative agreement with the theoretical arguments
proposed by Haehnelt, Natarajan \& Rees (1998) and Silk \& Rees
(1998).}

\end{itemize}

\acknowledgments I wish to thank Pat C\^ot\'e, David Merritt, Milos
Milosavljevic and Jerry Sellwood for providing the answers to many
questions. Comments from Uros Seljak, James Bullock and the anonymous
referee have also been of great help. This work is partly supported
by NASA/LTSA grant NAG5-8693, and has made use of the NASA/IPAC
Extragalactic  Database (NED).

\centerline{\includegraphics{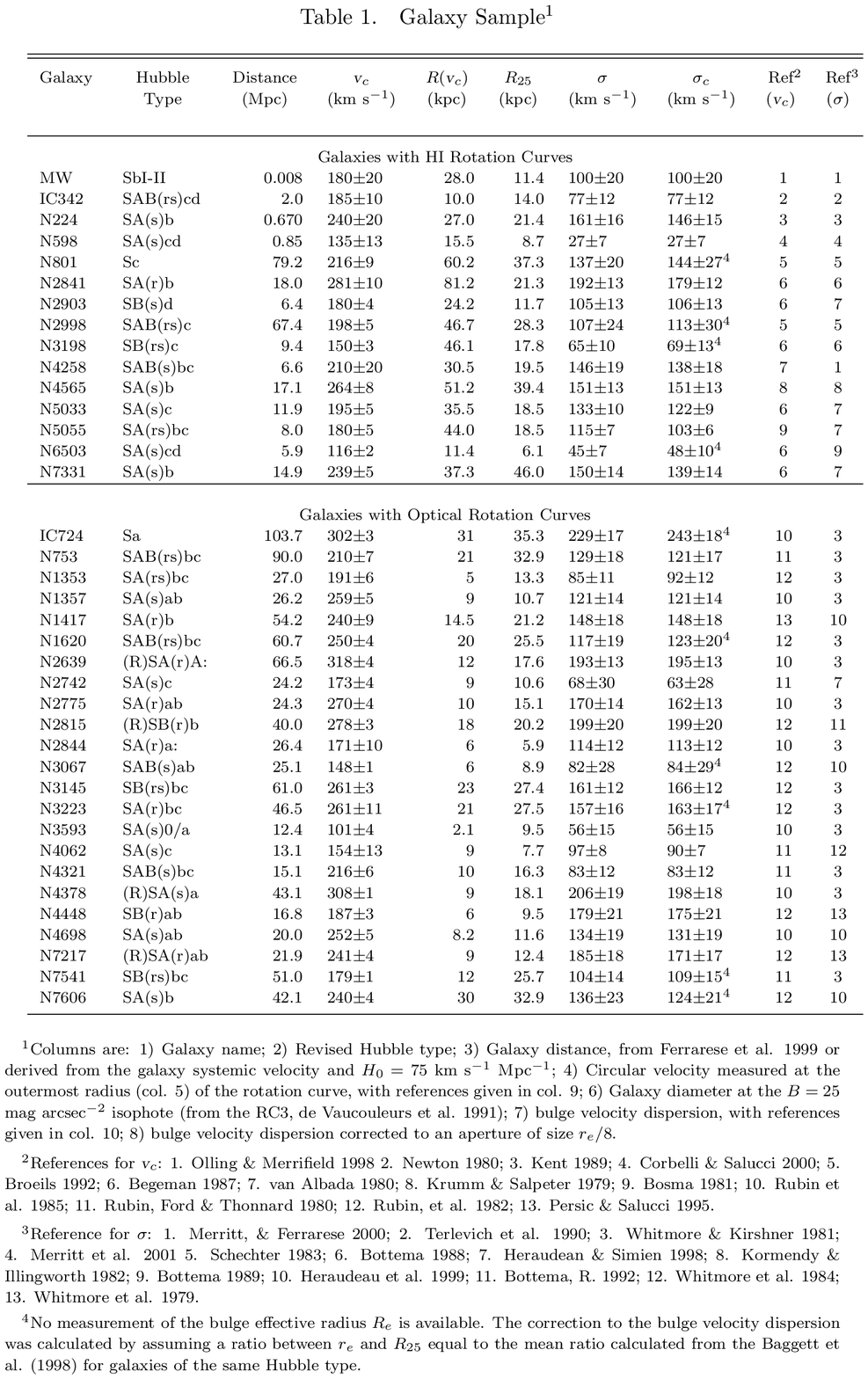}}  \centerline{}

\begin{deluxetable}{llrrl}
\tablecolumns{6} 
\tablewidth{0pc} 
\scriptsize \tablenum{2}
\tablecaption{Results of the Linear Regression Fits: $\vc = \alpha \sigma_c +
\beta$\label{tbl-2}} 
\tablehead{ 
\colhead{Sample\tablenotemark{(2)}} &
\colhead{$N$} & \colhead{$\alpha$} & 
\colhead{$\beta$} &
\colhead{$\chi_r^2$} } 
\startdata 
1. All spirals     			&  37 & 0.781$\pm$0.061  &  0.70$\pm$0.13 &  1.9  \nl     
2. Spirals, $R(\vc)/R_{25} > 1.0$  only &  15 & 0.683$\pm$0.060  &  0.89$\pm$0.12 &  0.70 \nl     
3. Spirals, $R(\vc)/R_{25} > 1.0$ and   &  13 & 0.839$\pm$0.089  &  0.55$\pm$0.19 &  0.38 \nl     
   \hskip .2in excluding N3198 and N6503&     &    		 &   		  &       \nl 
4. All  Ellipticals   			&  20 & 0.94$\pm$0.11    &  0.31$\pm$0.26 &  0.66 \nl
5. Samples 3 and 4 combined		&  33 & 0.892$\pm$0.041  &  0.44$\pm$0.09 &  0.51 \nl
\enddata
\tablenotetext{1}{Units for $\sigma_c$ and $\vc$ are km s$^{-1}$.}  
\tablenotetext{2}{M33 is excluded from all fits.}
\end{deluxetable}


\begin{references}

Adams, F.C., Graff, D.S., \& Richstone, D. 2000, ApJ, 551, L31

Akritas, M. G. \& Bershady, M. A. 1996, ApJ, 470, 706

Arnaboldi, M., et al. 1998, ApJ, 507, 759

Awaki, H., et al. 1994, PASJ, 46, L65

Baggett, W.E., Baggett, S.M., \& Anderson, K.S.J. 1998, AJ, 116, 1626

Begeman, K.G. 1987, PhD Thesis

Biermann, P.L., Kronberg, P.P., \& Schmutzler, T. 1989, 208, 22

Bosma, A. 1981, AJ, 86, 1825

Bottema R., 1988 A\&A 197, 105

Bottema, R. 1992, A\&A 257 ,69

Broeils 1992, PhD. Thesis

Bullock, J.S., et al. 2001, MNRAS, 321, 559

Casertano, S., \& van Gorkom, J.H. 1991, AJ, 101, 1231

Cattaneo, A., Haehnelt, M. G. \& Rees, M. J., 1999, MNRAS, 308, 77

C\^ot\'e, P. et al. 2001, ApJ, 559, 828

Corbelli, E., \& Salucci, P. 2000, MNRAS, 311, 441

Davies, M. et al. 1987, ApJ, 64, 581

de Blok, W.J.G. \& Bosma, A. 2002, A\&A, 385, 816

de Blok, W.J.G., McGaugh, S.S. \& Rubin, V. 2001, AJ, 122, 2396

de Vaucouleurs, G., de Vaucouleurs, A., Corwin, H. G., Buta, R. J.,
Paturel, G. \& Fouque, P. 1991, Third Reference Catalog of Bright
Galaxies (New York: Springer) (RC3)

Dressler, A. 1984, ApJ, 286, 97

Faber, S., \& Gallagher, J.S. 1979, ARAA, 17, 135

Fabricant, D., Rybicki, G., \& Gorenstein, P. 1984, ApJ, 286, 186

Ferrarese, L. 2002, to appear in {\em Current High Energy Emission
Around Black Holes}, eds. C.$-$H. Lee (astro-ph/0203047).

Ferrarese, L. \& Merritt, D. 2000, ApJ, 539, L9


Ferrarese, L., et al. 2000, ApJ, 529, 745

Gebhardt, K., et al. 2000, AJ, 119, 1157

Gebhardt, K., et al. 2001, AJ, 122, 2469

Gebhardt, K., et al. 2000, AJ, 119, 1157

Genzel, R., et al. 2000, astroph/0001428

Gerhard, O., Kronawitter, A., Saglia, R.P., \& Bender, R. 2001, AJ,
121, 1936

Graham, A.W., Erwin, P., Caon, N., \& Trujillo, I. 2001, ApJ, 563, L11

Haehnelt, M. G. \& Kauffmann, G., 2000, MNRAS, 318, L35

Haehnelt, M. G., Natarajan, P. \& Rees, M. J. 1998, MNRAS, 300, 817

Haehnelt, M. G., \& Rees, M.J. 1993, MNRAS, 263, 168

Irwin, J.A., \& Seaquist E.R. 1991, ApJ, 371, 111

Jorgensen, I., Franx, M. \& Kjaergaard, P. 1995, MNRAS, 276, 1341

Kent, S.M. 1987, in ``Nearly Normal Galaxies: from the Plank Time to
the Present'', eds. S.M. Faber (Springer, NY), p.81

Kent, S.M. 1989, PASP, 101, 489

Klypin, A., Zhao, H, \& Somerville, R.S., 2001, ApJ, submitted
(astro-ph/0110390)

Koribalski, B., Dickey, J.M., \& Mebold, U. 1993, ApJ, 402, L41

Kormendy, J., \& Illingworth, G. 1982, ApJ, 256, 460

Kormendy, J., \& Illingworth, G. 1983, ApJ, 265, 632

Kormendy, J., \& Richstone, D., 1995, ARAA, 27, 235

Kronawitter, A., Saglia, R.P., Gerhard, O., \& Bender, R. 2000, A\&AS,
144, 53

Krumm, N., \& Salpeter, E.E. 1979, AJ, 84, 1138

Lauer, T. 1985, ApJ, 292, 292

Loeb, A., \& Rasio, F. 1994, ApJ, 432, L52

Martini, P., \& Weinberg, D.H. 2001, ApJ, 437, 12

Matsushita, K., et al. 1998, ApJ, 499, 13

McElroy, D.B. 1995, ApJS, 100, 105

Merritt, D., \& Tremblay, B. 1993, AJ, 106, 2229

Merritt, D. \& Ferrarese, L. 2001b, MNRAS, 320, L30

Merritt, D. \& Ferrarese, L. 2001a, ApJ, 547, 140

Merritt, D., Ferrarese, L. \& Joseph, C. 2001, Science, 293, 1116

Merritt, D., \& Ferrarese, L. 2002 to appear in  {\em The Central Kpc
of Starbursts and AGNs}, eds. J.H. Knapen,  J.E. Beckman, I. Shlosman,
\& T.J. Mahoney (astro-ph/0107134)

Monaco, P., Salucci, P., \& Danese, L. 2000, MNRAS, 311, 279

Napolitano, N.R., Arnaboldi, M., Freeman, K.C., \& Capaccioli,
M. 2001, A\&A, in press (astro-ph/0109112)

Navarro, J.F., Frenk, C.S., \& White, S.D.M 1995, MNRAS, 275, 56

Navarro, J.F., \& Steinmetz, M. 2000, ApJ, 538, 477

Newton, K. 1980, MNRAS, 191, 169

Norman, C.A., Sellwood, J.A., \& Hasan, H. 1996, ApJ, 462, 114

Olling, R.P., \& Merrifield, M.R. 1998, MNRAS, 297, 943

Persic, M, \& Salucci, P. 1995, ApJS, 99, 501

Persic, M., Salucci, P., \& Stel, F. 1996, MNRAS, 281, 27

Robinson, I. 1965, {\em Quasi-Stellar Source and Gravitational
Collapse},  University of Chicago press

Rogstad, D.H., \& Shostak, G.S. 1972, ApJ, 220, L37

Rots, A.H., \& Shane, W.W. 1975; A\&A, 45, 25

Rubin, V., Ford, W.K., \& Thonnard, N. 1978, ApJ, 225, L107

Rubin, V., Ford, W.K., \& Thonnard, N. 1980, ApJ, 238, 471

Rubin, V., Thonnard N., \& Ford W.K. 1977, ApJ, 217, L1

Rubin, V. et al. 1982, ApJ, 261, 439

Rubin, V. et al. 1985, ApJ, 289, 81

Rupen M.P. 1991, AJ, 102, 48

Saglia, R.P., Kronawitter, A., Gerhard, O., \& Bender, R. 2000, AJ,
119, 153

Saika, D.J., et al. 1980, MNRAS, 245, 397

Schechter, P.L. 1983, ApJS 52, 425

Seljak, U. 2002, MNRAS, in press (astro-ph/0201450)

Silk, J. \& Rees, M. J. 1998, A\&A, 331, L4

Sofue, Y. \& Rubin, V. 2001, ARA\&A, 39, 137

Sofue, Y. et al. 1999, ApJ, 523, 136

Steinmetz, M., \& Muller, E. 1995, MNRAS, 276, 549

Terlevich, E., Diaz, A. I., Terlevich, R. 1990, MNRAS 242, 271

Umemura, M., Loeb, A., \& Turner, E.L. 1993, ApJ, 419, 459

van Albada, T.S. 1980, A\&A, 90, 123

van den Marel, R. 1991, MNRAS, 253, 710

van den Bosch, F. 2000, ApJ, 530, 177

van den Bosch, F. et al. 2000, AJ, 119, 1579

van der Kruit, P.C. 1974, ApJ, 188, 3

Zhang, B., \& Wyse, R., 2000, MNRAS, 313, 310

Whitmore, B.C., Kirshner, R.P. 1981, ApJ 250, 43

Whitmore, B.C., Kirshner, R.P., Schechter, P. L. 1979, ApJ 234, 68

Whitmore, B.C., Rubin, V.C., Ford, W.K., 1984 ApJ 287, 66

Yu, Q., \& Tremaine, S. 2002, MNRAS, in press.



\end{references}
\end{document}